\documentstyle[12pt,fleqn]{article}
\begin{document}
\setlength{\unitlength}{1mm}
\title{{\hfill {\small Alberta-Thy 09-96} } \vspace*{2cm}
\\
Black Hole Entropy: Thermodynamics,
Statistical-Mechanics and Subtraction Procedure
}
\author{\\
V.P.Frolov$^{1,2,3}$, D.V.
Fursaev$^{1,4}$, and
A.I.Zelnikov$^{1,3}$ \date{}}
\maketitle
\noindent  {
$^{1}${ \em
Theoretical Physics Institute, Department of Physics, \ University of
Alberta, \\ Edmonton, Canada T6G 2J1}
\\ $^{2}${\em CIAR Cosmology Program}
\\ $^{3}${\em P.N.Lebedev Physics Institute,  Leninskii Prospect 53,  
Moscow
117924, Russia}
\\ $^{4}${\em Joint Institute for
Nuclear Research, Laboratory of Theoretical Physics, \\
141 980 Dubna, Russia}
\\
\\
e-mails: frolov, dfursaev, zelnikov@phys.ualberta.ca
}
\bigskip

\begin{abstract}
The thermodynamical one-loop entropy $S^{TD}$ of a two-dimensional  
black hole
in thermal equilibrium with the massless quantum gas is calculated.  
It is shown
that $S^{TD}$ includes the Bekenstein-Hawking entropy, evaluated for  
the
quantum corrected geometry, and the finite difference of statistical  
mechanical
entropies $-Tr\hat{\rho}\ln\hat{\rho}$ for the gas on the black hole  
and
Rindler spaces. This result demonstrates in an explicit form that the  
relation
between thermodynamical and statistical-mechanical entropies of a  
black hole
is non-trivial and requires special subtraction procedure.
\end{abstract}

\bigskip

{\it PACS number(s): 04.60.+n, 12.25.+e, 97.60.Lf, 11.10.Gh}

\baselineskip=.6cm

\newpage

\noindent
\section{Introduction}

Black holes are known to behave as usual thermodynamical systems with  
the
entropy
	\begin{equation}\label{1.1}
	{S}^{BH} ={ 1 \over 4 } {A^H \over  
l_{\mbox{\scriptsize{P}}}^{~2}},
	\end{equation}
where ${A}^H$   is   the   area   of the event  horizon
\cite{Beke:72},\cite{Hawk:75}. The Bekenstein-Hawking entropy  
${S}^{BH}$ is the
physical quantity which can be measured  in (gedanken) experiments by  
making
use of
the first law of thermodynamics which can be represented in the form
\cite{GiHa:76}:
	\begin{equation}\label{1.2}
	dF^{H}=-{S}^{BH}dT^{H}~~~.
	\end{equation}
Here $M^{H}$ is mass of a black hole, $T^{H}=(8\pi M^{H})^{-1}$ is  
its Hawking
temperature and $F^{H}=M^{H}-T^{H}S^{BH}$ is the free energy.

The fundamental aspect of the black hole thermodynamics is its
statistical-mechanical foundation. This implies the solution of the  
following
problems: (1) definition of internal degrees of freedom of a black
hole and finding the corresponding density matrix $\hat{\rho}^H$;
(2) calculation of the statistical-mechanical entropy
$S^{SM}=-\mbox{Tr}(\hat{\rho}^H\ln \hat{\rho}^H )$;
(3) establishing the relation between
$S^{SM}$ and the observable thermodynamical $S^{TD}$ entropy defined  
by the
first law (\ref{1.2}).

There is a number of alternatives in choosing internal degrees of  
freedom of a
black hole that might be responsible for its entropy. One of the  
promising
suggestions is to relate these degrees of freedom with thermal  
excitations of
quantum fields around a black hole \cite{Hoof:85}.  However the  
explicit
calculations give an infinite value for the corresponding entropy
$S^{SM}=-\mbox{Tr}(\hat{\rho}^H\ln \hat{\rho}^H )$. This fact is in  
obvious
contradiction with the finiteness of the observable quantity  
(\ref{1.1}).  The
same problem appears in the attempts to interpret $S^{BH}$ as an  
entanglement
entropy connected to the loss of information about correlations of  
quantum
states inside and outside the horizon \cite{Sorkin}. This  
contradiction
indicates that the relation between statistical mechanics and  
thermodynamics in
the case of black holes may be non-trivial.
The property which singles out black holes from other thermodynamical  
systems
is that the Hamiltonian describing the quantum fields on a black hole
background depends on the mass $M^{H}$ of a black hole and, hence, on  
the
equilibrium temperature $T^H$. This has two consequences. i) The  
statistical
entropy $S^{SM}$ does not coincide with the thermodynamical entropy  
of a black
hole \cite{Frol:95}. ii) The calculation of $S^{SM}$ requires  
off-shell methods
when mass of a black hole and temperature of the system are  
considered as
independent parameters.

In this Letter we compare two frequently used off-shell procedures  
(the so
called conical singularity method and brick wall approach) and obtain  
relation
between them for 2D quantum black hole models.
This analysis enables us to establish the explicit structure of the  
total
one-loop black hole entropy $S^{TD}$.
It is shown that $S^{TD}$ is related to the quantity obtained from  
entropy
$S^{SM}=-\mbox{Tr}(\hat{\rho}^H\ln \hat{\rho}^H )$ by subtraction   
the
statistical-mechanical entropy for some "reference" Rindler-like  
background.

\section{On-shell and off-shell approaches}
\setcounter{equation}{0}

Thermodynamical characteristics of a black hole are defined by the  
partition
function
	\begin{equation}\label{2.1}
	Z=\int [D\phi] e^{-I[\phi]}~~~.
	\end{equation}
Here $I[\phi]$ is the Euclidean classical action and all the physical  
variables
$\phi$ including the metric $\gamma_{\mu\nu}$ are assumed to be  
periodic in the
Euclidean time $\tau$ with the period $\beta_\infty$. We assume that  
metrics
are asymptotically flat.  We confine our analysis to a simple quantum  
model
described by the action
	\begin{equation}\label{2.2}
	I[\phi]=I_{grav}+\frac 12 \int
	\sqrt{\gamma}\varphi_{,\mu}\varphi^{,\mu}d^2x~~~
	\end{equation}
where $\varphi$ is the massless matter field and $I_{grav}$ is the  
part of the
action which includes the gravitational degrees of freedom. Although  
the
results  do not depend on a particular form of $I_{grav}$,
it is convenient to take as an example the model of two-dimensional  
dilaton
gravity
	\begin{equation}\label{2.3}
	I_{grav}[\gamma_{\mu\nu},r]=-{1 \over 4} \int_{M^2}^{}(r^2  
R+2(\nabla
r)^2+2)\sqrt{\gamma}d^2x
 	-{1 \over 2}\int_{\partial M^2}^{}r^2 (k-k_0)dy ~~~.
	\end{equation}
The action (\ref{2.3}) can be obtained by spherical symmetric  
reduction of the
four-dimensional Einstein gravity, the radius $r$ being the dynamical  
variable
playing the role of the dilaton field. To get a well defined  
canonical ensemble
we also suppose that the black hole is in the spherical cavity of a  
radius
$r_B$ (see \cite{York:86}). The inverse temperature measured at $r_B$  
is
denoted by $\beta$.

To evaluate the integral (\ref{2.1}) we will work under the  
assumption that
only matter field $\varphi $ is quantized while the gravitational  
variables
$\gamma_{\mu\nu}$ and $r$ are taken into account in the  
quasiclassical
approximation
	\begin{equation}\label{2.4}
	\left. {\delta I / {\delta \phi}}\right| _{\phi=\phi_0}  
=0~~~.
	\end{equation}
Then the partition function (\ref{2.1}) can be expressed
in terms of the effective action $W(\beta)$
	\begin{equation}\label{2.6}
	\ln Z(\beta)=                
W(\beta)=I[\phi_0(\beta)]+W_1[\phi_0(\beta)]~~~,
	\end{equation}
	\begin{equation}\label{2.7}
	W_1[\phi_0(\beta)]=\frac 12 \log \det  
\nabla^{\mu}\nabla_{\mu}~~~.
	\end{equation}
Here $\phi_0=\{\gamma^{(0)}_{\mu\nu}, \varphi=0\}$ is the classical  
extremum of
the action (\ref{2.2}) in the black hole sector. For	the dilaton  
model
(\ref{2.3}) the classical metric has the Schwarzschild form
	\begin{equation}\label{2.8}
	ds^2=\gamma^{(0)}_{\mu\nu}~dx^\mu
	dx^\nu=(1-r_+/ r)d\tau^2+ (1-r_+/ r)^{-1}dr^2~~~, 	 
\end{equation}
$0\leq \tau\leq 4\pi r_+=T_H^{-1}~$. Functional $W_1$ is the one-loop  
quantum
correction.  According to a standard procedure the ultraviolet  
divergences in
$W_1$ should be removed by the renormalization of  bare cosmological  
and
gravitational constants in the classical gravitational action  
(\ref{2.2}). In
order to preserve the asymptotical flatness we put the renormalized
cosmological constant to be $-1/2$. The term which is an integral of  
a
curvature can be omitted  because in two dimensions it is a  
topological
invariant.

In thermal equilibrium the system is uniquely described by fixing two  
external
parameters: the inverse temperature $\beta$ on the boundary and the  
'radius'
$r_B$. All other characteristics such as the 'radius' of the event  
horizon
$r_+$ are
the functions of $\beta$ and $r_B$. For instance, for the dilaton  
model
(\ref{2.3}) $r_+$ is determined by the equation
$\beta=4\pi r_+(1-r_+/r_B)^{1/2}$.

The {\it thermodynamical entropy}  of a black hole $S^{TD}$ is  
defined by  the
response of the  free energy $F(\beta) =\beta^{-1}W(\beta)$ to the  
change of
the inverse temperature $\beta$ for fixed $r_B$.
	\begin{equation}\label{2.11}
	S^{TD}(\beta)=\beta^{2}  {d F(\beta)\over d\beta}  
=\left(\beta {d \over
	d\beta}-1\right)W(\beta)=S_0^{TD}+S_1^{TD}~~~.
	\end{equation}
It can be shown \cite{York:86} that
	\begin{equation}\label{2.13}
	S_0^{TD}[\phi_0(\beta)]=\left(\beta {d \over d\beta}
-1\right)I[\phi_0(\beta)]~~~,
	\end{equation}
coincides with the Bekenstein-Hawking entropy $S^{BH}$ given by  
Eq.(\ref{1.1}),
while $S^{TD}_1(\beta)$ obtained from $W_1[\phi_0(\beta)]$
describes the quantum correction. This correction contains also the
entropy of the thermal radiation outside the black hole as its part.  
By its
construction the thermodynamical entropy $S^{TD}$  is well defined  
and finite.
All the calculations required to obtain this quantity can be  
performed {\em
on-shell}, that is on  regular complete vacuum  solutions of the  
Euclidean
gravitational equations.

Eq.(\ref{2.11}) contains the renormalized effective action $W$
calculated on a particular classical solution. This renormalized  
action itself
is defined as a functional
$
W[\phi]=I[\phi]+W_1[\phi]
$
for an arbitrary field $\phi$ with appropriately chosen boundary  
conditions.
The extremum $\bar{\phi}(\beta)$ of this functional
	\begin{equation}\label{2.16}
	\left.
	{\delta W\over{\delta \phi}}\right| _{\phi=\bar{\phi}(\beta)}  
=0
	\end{equation}
describes a modified field configuration which differs from a
classical solution by quantum corrections:
$\bar{\phi}(\beta)=\phi_0(\beta) +\hbar\phi_1(\beta)$.
The important observation is that, if one is interested in
one-loop effects, the difference between the values of $W$ on  
$\phi_0$ and
$\bar{\phi}$ turns out to be of the second order
in  Planck constant $\hbar$
	\begin{equation}\label{2.17}
	W(\beta)=W[\phi_0(\beta)]=W[\bar{\phi}(\beta)]+  
O(\hbar^2)~~~.
	\end{equation}
This follows from (\ref{2.16}), provided the quantum corrected and  
classical
solutions obey the same boundary conditions.

As we discussed earlier one must use off-shell methods to find out  
the
statistical-mechanical interpretation of the black hole entropy
(\ref{2.11}). In off-shell approaches the mass of a black hole is  
considered as
an additional parameter, independent from the temperature which is  
associated
with the periodicity of the Euclidean manifold.
However, it is well known that such manifolds are not regular on the  
Euclidean
horizon.

We discuss and compare here two off-shell approaches frequently used  
for the
calculation of the entropy. The first method (conical singularity  
method) is to
work directly with the complete singular instanton \cite{SU}. Another  
procedure
(brick wall model) is to introduce at some small proper distance  
$\epsilon$
from the black hole horizon an additional mirror like boundary  
\cite{Hoof:85}
and to work with incomplete manifold. The brick wall model has a  
clear
statistical-mechanical interpretation, while the conical-singularity  
method
enables one to relate thermodynamical and off-shell entropies.

The one-loop entropy $S^{CS}$ evaluated in the conical singularity  
method is
defined from the effective action as
	\begin{equation}\label{2.18}
	S^{CS}(\beta)
	=\left(\beta {\partial \over \partial
 	\beta}-1\right)W[\bar{\phi},\beta]~~~.
	\end{equation}
The derivative over $\beta$ in (\ref{2.18}) is taken when the  
solution
$\bar{\phi}$ (as well as the cavity radius $r_B$) is fixed. One  
should put the
on-shell condition $\bar{\phi}=
\bar{\phi}(\beta)$ only after the differentiation. The entropy  
$S^{CS}$ can be
represented as the sum
	\begin{equation}\label{2.19}
	S^{CS}(\beta)=S_0^{CS}[\bar{\phi}(\beta)]+S_1^{CS}[\bar{\phi}( 
\beta)]
	\end{equation}
of the tree-level part
	\begin{equation}\label{2.20}
	S_0^{CS}[\bar{\phi}(\beta)]=\left(\beta {\partial \over  
\partial
\beta}-1\right)I[\bar{\phi},\beta]~~~,
	\end{equation}
and a quantum correction to it
	\begin{equation}\label{2.21}
	S^{CS}_1(\beta)=\left(\beta {\partial \over \partial
\beta}-1\right)W_1^{CS}[\bar{\phi},\beta]~~~.
	\end{equation}
The tree-level contribution $S^{CS}_0$, defined by the classical  
action on the
manifold with conical singularity is \cite{FS1}
	\begin{equation}\label{2.22}
	S_0^{CS}[\bar{\phi}(\beta)]=\bar{S}^{BH} ={ 1 \over 4 }  
\bar{A}^H~~~.
	\end{equation}
It has the form of the Bekenstein-Hawking entropy with the area $A^H$  
of the
classical horizon replaced by the horizon area $\bar{A}^H$ of the  
quantum
corrected solution.
For the dilaton gravity (\ref{2.3}) the latter is
$S_0^{CS}[\bar{\phi}(\beta)]=\pi \bar{r}_+^2$
where $\bar{r}_+$ is the value of the dilaton at the black hole  
horizon.

Remarkably, when $\bar{\phi}=
\bar{\phi}(\beta)$ the total off-shell one-loop entropy (\ref{2.18})  
coincides
with the total thermodynamical (on-shell) entropy (\ref{2.11})
	\begin{equation}\label{2.24}
	S^{CS}(\beta)=S^{TD}(\beta)~~~.
	\end{equation}
To obtain this relation one must take into account Eq.(\ref{2.17})  
for the
on-shell effective action and rewrite the total derivative in  
(\ref{2.11}) in
the form
	\begin{equation}\label{2.25}
	{d W[\bar{\phi}(\beta)]\over d\beta}= {\partial  
W[\bar{\phi},\beta]\over
\partial 	\beta}+\left.\int_{M^2}{\delta W[\phi,\beta] \over  
\delta
\phi}\right|_{\phi=\bar{\phi}(\beta)}
	{\partial \bar{\phi}(\beta) \over \partial \beta}~~~.
	\end{equation}
The partial derivative with respect to $\beta$ in the first term in  
the right
hand side of (\ref{2.20}) is taken at the fixed solution $\bar{\phi}$  
(at the
fixed black hole mass) and
the last term in this equation vanishes
because $\bar{\phi}(\beta)$ is the extremum of the effective action.
The variational procedure in (\ref{2.25}) involves
the generalized class of spaces including ones with conical  
singularities.
However in our case the extremum of $W$ remains at the smooth  
instanton
$\bar{\phi}(\beta)$. The extrema on  manifolds with conical  
singularities are
ruled out in the absence of a matter singularly distributed over the  
horizon.

Eqs. (\ref{2.24}), (\ref{2.19}) and (\ref{2.22}) enable one to  
represent the
one-loop thermodynamical entropy of a quantum black hole in the  
following form
	\begin{equation}\label{2.26}
	S^{TD}(\beta)=\bar{S}^{BH}(\beta)+S_1^{CS}(\beta)~~~.
	\end{equation}
The last term $S_1^{CS}(\beta)$ in (\ref{2.26}) is evaluated from the  
quantum
determinant (\ref{2.7}) on the manifold with conical singularities.
For the two-dimensional model (\ref{2.2}) where quantum effects are  
generated
by the massless scalar field $\varphi$ it can be found explicitly.

To do that note the conical singularities do not result in the new  
divergences
in the black hole entropy \cite{SU}.
So we can define $W_1^{CS}$ as the renormalized value of the  
determinant
(\ref{2.7}) on a two-dimensional manifold with the conical  
singularity. The
metric on this manifold can be written in the form (\ref{2.8})
	\begin{equation}\label{2.27}
	ds^2=g(r)d\tau ^2+g^{-1}(r)dr^2~~~,
	\end{equation}
where $0\leq \tau \leq \beta_\infty$ and $r_h\leq r \leq r_B$. The  
conical
angle deficit $2\pi(1-\alpha)$ near the horizon $r=r_h$ is  
parametrized by
$
\alpha=\beta_\infty /\beta_H=\beta /(\beta_H \sqrt{g(r_B)}),
$
and $\beta_H^{-1}=g'/4\pi$ is defined by the surface gravity.
To find the entropy correction $S_1^{CS}$ it is convenient to map
the singular instanton (\ref{2.27}) onto a cone $C_\alpha$ with the  
line
element
	\begin{equation}\label{2.29}
	d\tilde{s}^2=y^2d\tilde{\tau}^2+dy^2=\mu^{-2}e^{-2\sigma}ds^2
	\end{equation}
where $0\leq \tilde{\tau} \leq 2\pi \alpha$ and $0\leq y \leq 1$.  
($\mu$ is an
arbitrary parameter with dimension (length)$^{-1}$.)
The map (\ref{2.29}) is explicitly described by the formulas
	\begin{equation}\label{2.30}
	y(r)=\exp\left({2\pi \over \beta_\infty}\int_{r_B}^{r}{dz  
\over
g(z)}\right)~~~,
	\end{equation}
	\begin{equation}\label{2.31}
	\sigma(r)=\frac 12 \int_{r_B}^{r}{dz \over g}\left(g'-{4\pi  
\over
\beta_H}\right)+\ln\left({\beta_H \sqrt{g(r_B)} \over  
2\pi\mu}\right)~~~.
	\end{equation}
The actions $W_1^{CS}$ on the instanton (\ref{2.27}) and on the cone
(\ref{2.29}) are related via the anomalous conformal transformation  
\cite{FFZ}
	\begin{eqnarray}
	W_1^{CS}[\gamma_{\mu\nu},\beta,r_B]&=&W_1[C_\alpha]-{1 \over
24\pi}\left[\int_{M^2}
	\left(R\sigma-(\nabla \sigma)^2\right) +\int_{\partial  
M^2}\left(2k \sigma
+3\sigma_{,\mu}n^{\mu}\right)\right]  \nonumber \\
	&+& {1 \over 12}\left(\alpha -{1 \over  
\alpha}\right)\sigma(r_h)~~~
	\label{2.32}
	\end{eqnarray}
where the last term is the contribution of the vertex point $r_h$ of  
the
singular instanton. The expression in the square brackets is  
proportional to
$\beta$ and hence does not contribute to $S_1^{CS}$. The equation  
(\ref{2.21})
applied to (\ref{2.32}) at $\alpha=1$ gives
	\begin{equation}\label{2.33}
	S_1^{CS}[\beta,r_B]=\frac 16 \sigma(r_h)~~~
	\end{equation}
where we omitted a numerical constant produced by $W_1[C_\alpha]$.  
Thus the
one-loop correction
evaluated by the conical singularity method for massless quantum  
field is
completely determined by the value $\sigma(r_h)$ of the conformal  
factor at the
horizon \cite{FS1},\cite{Myers}.

\section{Relation with statistical mechanics}
\setcounter{equation}{0}

The correction $S_1^{CS}$ to the black hole entropy computed by the
conical singularity method has a well defined geometrical meaning
but it does not coincide with a statistical-mechanical entropy
$-\mbox{Tr}(\hat{\rho}^H\ln \hat{\rho}^H)$ for any density matrix.  
Let us
establish now the relation of the conical singularity method with the  
brick
wall model which has the simple statistical-mechanical  
interpretation.
For this purpose we introduce at some small proper distance  
$\epsilon$ from the
black hole horizon an additional mirror like boundary with the  
Dirichlet
condition on it \cite{Hoof:85}.
We denote the corresponding manifold by $M^2_\epsilon$.
In this model the horizon does not belong to a manifold and it is  
possible to
formulate the canonical ensemble for the quantum field $\varphi$  
between two
mirrors at arbitrary temperature $\beta^{-1}$ (measured as before at  
the
external boundary). Then all the information about  thermodynamics of  
the
system can be derived from the free energy ${\cal F}(\beta)$
	\begin{equation}\label{3.1}
	e^{-\beta{\cal F}(\beta)}=\mbox{Tr}(e^{-\beta \hat{H}})~~~,
	\end{equation}
where $\hat{H}=\hat{H}(\gamma_{\mu\nu})$ is the Hamiltonian for field
$\varphi$. The free energy ${\cal F}(\beta)$ can be computed if the  
spectrum of
$\hat{H}$ is known. The alternative way is a field-theoretical  
computation
which enables to relate this quantity to the effective action  
$W_1^{BW}$ for
$\varphi$ on a static manifold with two boundaries
	\begin{equation}\label{3.2}
	W_1^{BW}(\beta)=\beta {\cal F}(\beta) +\beta U_H + \beta  
U_A~~~.
	\end{equation}
The quantities $U_H$ and $U_A$ in the right hand side of (\ref{3.2})  
do not
depend on $\beta$ and represent a contribution of the vacuum energy.  
The first
quantity $\beta U_H$ originates from the difference between the  
covariant
measure $\prod_{x}(\det\gamma(x))^{1/4}d\varphi(x)$ in the functional  
integral
(\ref{2.1}) and the Hamiltonian measure
$\prod_{x}(\gamma_{00}(x))^{-1/2}(\det\gamma(x))^{1/4}d\varphi(x)$  
used in the
canonical formalism for the path integral representation for
${\cal F}(\beta)$ (see \cite{DeAlwis}). The second term $\beta U_A$  
is finite
and accounts for,
according to Allen \cite{Alle:86}, a difference of the renormalized  
covariant
path integral and the partition function
(\ref{3.1}) defined by the spectrum of the normally ordered  
Hamiltonian.

The density matrix for the brick wall canonical ensemble is
	\begin{equation}\label{3.3}
	\hat{\rho}^H_\epsilon(\beta)={e^{-\beta \hat{H}} \over
	\mbox{Tr}(e^{-\beta \hat{H}}) }~~~,
	\end{equation}
where superscript $H$ indicates that field $\varphi$ is given on the  
black hole
background and subscript $\epsilon$ indicates the position of the  
internal
boundary. Thus the entropy for the field
in  the brick-wall model reads
	\begin{equation}\label{3.4}
	S_1^{BW}[\beta,\epsilon,r_B]=-\mbox{Tr}(\hat{\rho}^H_\epsilon( 
\beta)\log
\hat{\rho}^H_\epsilon(\beta))=
	\beta^2{\partial \over \partial \beta}{\cal F}(\beta)~~~,
	\end{equation}
and using  Eq. (\ref{3.2}) it can be rewritten in the  
field-theoretical form
	\begin{equation}\label{3.5}
	S_1^{BW}[\beta,\epsilon,r_B]=\left(\beta {\partial \over  
\partial
\beta}-1\right)W_1^{BW}[\beta,\epsilon,r_B]~~~.
	\end{equation}
Using Eq.(\ref{3.5}) it is easy now to compute $S_1^{BW}$ and compare  
it with
the entropy $S_1^{CS}$ in the conical
singularity method. For this aim it is convenient to map conformally  
the
brick-wall space $M^2_\epsilon$ onto the cylinder  
$Q_{\alpha,\epsilon_z}$ with
the metric
	\begin{equation}\label{3.6}
	d\bar{s}^2=(d\tilde{\tau}^2+dz^2)=\mu^{-2}y^{-2}e^{-2\sigma}ds 
^2
	\end{equation}
where $0\leq \tilde{\tau} \leq 2\pi \alpha$, $ds^2$ is the black hole  
metric
(\ref{2.27}), $y$ and $\sigma$ are given by (\ref{2.30}) and  
(\ref{2.31}). The
coordinate $z=-\ln y$ on cylinder ranges
in the interval $0\leq z \leq \epsilon_z$, where  $\epsilon_z$ is  
related with
the proper distance $\epsilon$ from the "brick-wall" to the horizon  
(in the
limit $\epsilon\rightarrow 0$) as
	\begin{equation}\label{3.7}
	\epsilon_z=-\ln{\epsilon \over \mu} + \sigma(r_h)~~~.
	\end{equation}
The effective actions $W_1^{BW}[\beta,\epsilon,r_B]$ on the brick  
wall space
$M^2_\epsilon$ and the effective action $W_1[Q_{\alpha,\epsilon_z}]$  
on the
cylinder  differ by the conformal anomaly term. But because the  
spaces are
static and regular everywhere, this difference is proportional to  
$\beta$ and
it does not contribute to the entropy. So one can rewrite the  
expression
(\ref{3.5}) for $S_1^{BW}$ as
	\begin{equation}\label{3.8}
	S_1^{BW}[\beta,\epsilon,r_B]=\left(\alpha {\partial \over  
\partial
\alpha}-1\right)W_1[Q_{\alpha,\epsilon_z}]~~~.
	\end{equation}
Here we took into account that $\alpha$ is proportional to $\beta$.
The action $W_1[Q_{\alpha,\epsilon_z}]$ of massless field on a two
dimensional cylinder of a large size $\epsilon_z$ can be calculated  
exactly in
this limit \cite{FFZ}
	\begin{equation}\label{3.9}
	W_1[Q_{\alpha,\epsilon_z}]=-{1 \over 12\alpha}\epsilon_z  
-\frac 12 \ln {\pi
\alpha \over \epsilon _z} + o(\epsilon_z^{-1})~~~.
	\end{equation}
Thus, using (\ref{3.7}), (\ref{3.8}) and (\ref{3.8}) one gets
	\begin{equation}\label{3.10}
	S_1^{BW}[\beta,\epsilon,r_B]=\frac 16 \sigma(r_h) - \frac  
16\ln {\epsilon
\over 	\mu}+\frac 12\ln{\pi \over \ln(\mu/\epsilon)} +
	O(|\ln\epsilon|^{-1})~~~.
	\end{equation}
Comparing (\ref{3.10}) with Eq.(\ref{2.33}) we see that the
statistical-mechanical entropy computed in the brick wall model  
differs from
$S_1^{CS}$ by  terms divergent in the limit $\epsilon\rightarrow 0$.  
Note that
the dependence of the entropy on the parameter $\mu$ reflects an  
arbitrariness
in the definition of the entropy and it is physically unobservable.

We shall demonstrate now that the difference $~S_1^{CS}-S_1^{BW}~$  
allows (at
least in 2D case) a quite simple presentation, which makes  
transparent the
relation between these quantities.
Let us compare the obtained result (\ref{3.10}) with the calculation  
in the
Rindler space. Consider the metric
	\begin{equation}\label{3.11}
	ds^2_R=y^2d\tilde{\tau}^2+ dy^2
	\end{equation}
where $0\leq \tilde{\tau} \leq 2\pi \alpha$ and $\epsilon \leq y\leq  
\mu$. This
choice of the parameters corresponds to the situation when the field  
is located
between two mirrors
at proper distances $\epsilon$ and $\mu$ from the Rindler horizon  
respectively.
The inverse temperature of the system measured at the distance $\mu$  
is
$2\pi\alpha\mu$. The entropy of the massless scalar field in this  
space can be
computed by the same method as in the case of the black hole. The  
Rindler space
(\ref{3.11}) can be conformally mapped onto the cylindrical space  
(\ref{3.6}).
The
size of the cylinder now is $\epsilon^R_z=-\ln(\epsilon/\mu)$.
Caring out the calculations one gets the brick wall  entropy in the  
Rindler
space (at $\alpha=1$)
	\begin{equation}\label{3.12}
	(S_R)_1^{BW}[2\pi\mu,\epsilon, \mu]=- \frac 16\ln {\epsilon  
\over \mu}+\frac
12\ln{\pi \over \ln(\mu/\epsilon)} +
	O(|\ln\epsilon|^{-1})~~~.
	\end{equation}
Therefore, comparing (\ref{3.12}) and (\ref{3.10}), we get
	\begin{equation}\label{3.13}
	S_1^{BW}[\beta,\epsilon,r_B]=\frac 16
 	\sigma(r_h)+(S_R)_1^{BW}[2\pi\mu,\epsilon, \mu]~~~.
	\end{equation}
Now, taking into account (\ref{2.33}), one can conclude that the  
conical
singularity entropy can be identically rewritten as the difference of  
two
statistical-mechanical entropies
	\begin{equation}\label{3.14}
	S_1^{CS}[\beta,r_B]=S_1^{BW}[\beta,\epsilon,r_B]-(S_R)_1^{BW}
	[2\pi\mu,\epsilon, \mu]~~~.
	\end{equation}
Eventually, from (\ref{2.26}), (\ref{3.4}) and (\ref{3.14}) it  
follows that the
total one-loop entropy of a two-dimensional black hole is
	\begin{equation}\label{3.15}
	S^{TD}(\beta)=
	\bar{S}^{BH}(\beta)+
\left(-\mbox{Tr}(\hat{\rho}^H_\epsilon(\beta)\log
	\hat{\rho}^H_\epsilon(\beta)\right)-
	\left(-\mbox{Tr}(\hat{\rho}^R_\epsilon(2\pi\mu)\log
\hat{\rho}^H_\epsilon(2\pi\mu))\right)~~~
	\end{equation}
where $\hat{\rho}^H_\epsilon(2\pi\mu)$ is the density matrix for the  
Rindler
space. Eq. (\ref{3.15}) is the main result of our Letter.
The first term in the right hand side of Eq.(\ref{3.15}) is the
Bekenstein-Hawking entropy determined by the area of the event  
horizon of
quantum corrected black hole. The last two terms represent the finite
difference of two divergent statistical-mechanical entropies  
calculated for the
black hole and for a corresponding "reference" Rindler space,  
respectively.
It is worth mentioning
that the similar subtraction formula naturally arises in the membrane  
paradigm
\cite{ThPrMa:86}. Namely, in order to obtain the correct
expression for the flux of the entropy
onto a black hole, Thorne and Zurek \cite{ThPrMa:86,ZuTh:85}
proposed to subtract from the entropy, calculated by the  
statistical-mechanical
method, the entropy of a thermal atmosphere of the black hole. The  
latter
entropy close to the horizon coincides with $(S_R)_1^{BW}$. So  
Eq.(\ref{3.15})
can be used to prove this conjecture.

\section{Summary}

As we have demonstrated different off-shell procedures
(conical singularity method and brick wall model) result
in different renormalized one-loop corrections to the black hole  
entropy. The
quantity $S_1^{CS}$ is finite, while $S_1^{BW}$ diverges as
$\epsilon\rightarrow 0$.
The quantities obtained by both methods also have different  
interpretations.
The brick wall entropy $S_1^{BW}$ is pure statistical-mechanical  
entropy, while
the quantity $S_1^{CS}$
can be represented only as the difference of two such entropies. In  
two
dimensions for the massless scalar field their relation can be  
established
exactly. It is also worth mentioning that instead of the brick-wall  
method one
can use another way to calculate the statistical entropy. As it was  
shown in
\cite{FFZ} this does not change the final subtraction equation  
(\ref{3.14}).

Our analysis demonstrates in an explicit form that the black holes  
are some
peculiar systems where thermodynamical and statistical-mechanical  
entropies may
not coincide. The thermodynamical computations which operate with the  
system
being in thermal equilibrium (obeying the
equations of motion) always give the finite result for the entropy in  
terms of
the renormalized gravitational coupling constants. It is the  
thermodynamical
entropy that is observable
(at least in the gedanken experiments).
On the other hand, the statistical-mechanical black hole
entropy includes physical divergences because of the infinite blue  
shift at the
horizon. In general the relation between thermodynamics and  
statistical
mechanics for black holes may require  some subtraction formula  
similar
to Eq.(\ref{3.15}).

In our study we did not deal with the statistical explanation of the  
tree-level
Bekenstein-Hawking entropy $\bar{S}^{BH}(\beta)$ itself. In the total  
entropy
(\ref{3.15}) $\bar{S}^{BH}(\beta)$ appears as the entropy of a  
quantum
corrected black hole and it is completely defined by the geometry.
So one can expect that it is possible to get the  
statistical-mechanical
interpretation of the Bekenstein-Hawking entropy  
$\bar{S}^{BH}(\beta)$ in the
theory where the metric itself arises as the result of quantum  
effects. Models
of induced gravity and superstrings do possess this property.

\vspace{12pt}
{\bf Acknowledgements}:\ \ This work was supported  by the Natural  
Sciences and
Engineering
Research Council of Canada.

\end{document}